\documentclass[aps,prl,epsfig,floats,twocolumn,amssymb,amsmath,showpacs]{revtex4}

\usepackage{graphicx}
\usepackage{epstopdf}

\newcommand{\beq}{\begin{equation}}
\newcommand{\eeq}{\end{equation}}
\newcommand{\beqa}{\begin{eqnarray}}
\newcommand{\eeqa}{\end{eqnarray}}
\newcommand{\bem}{\begin{math}}
\newcommand{\eem}{\end{math}}

\newcommand{\nvec}{\hat{\bf n}}

\newcommand{\bfx}{{\bf x}}

\newcommand{\bfv}{{\bf v}}

\begin{document}
\title{Electrokinetic Effects in Catalytic Pt-Insulator Janus Swimmers}

\author{S. Ebbens}
\email{s.ebbens@sheffield.ac.uk}
\affiliation{Department of Chemical and Process Engineering, University of Sheffield,
Sheffield, S1 3JD, UK}

\author{D.A. Gregory}
\affiliation{Department of Chemical and Process Engineering, University of Sheffield,
Sheffield, S1 3JD, UK}

\author{G. Dunderdale}
\affiliation{Department of Chemical and Process Engineering, University of Sheffield,
Sheffield, S1 3JD, UK}

\author{J.R. Howse}
\affiliation{Department of Chemical and Process Engineering, University of Sheffield,
Sheffield, S1 3JD, UK}

\author{Y. Ibrahim}
\affiliation{School of Mathematics, University of Bristol, Clifton, Bristol BS8 1TW, UK}

\author{T.B. Liverpool}
\affiliation{School of Mathematics, University of Bristol, Clifton, Bristol BS8 1TW, UK}

\author{R. Golestanian}
\email{ramin.golestanian@physics.ox.ac.uk}
\affiliation{Rudolf Peierls Centre for Theoretical Physics, University of Oxford, Oxford OX1 3NP, UK}

\date{\today}

\begin{abstract}
The effect of added salt on the propulsion of Janus platinum-polystyrene colloids in hydrogen peroxide solution is studied experimentally.
It is found that micromolar quantities of potassium and silver nitrate salts reduce the swimming velocity by similar amounts, while leading to significantly
different effects on the overall rate of catalytic breakdown of hydrogen peroxide. It is argued that the seemingly paradoxical experimental
observations could be theoretically explained by using a generalised reaction scheme that involves charged intermediates and has the topology of two nested loops.
\end{abstract}

\maketitle

\paragraph{Introduction.} In recent years there has been a flurry of activity in developing micro- and nanoscale self-propelling devices that are engineered to produce enhanced motion within a fluid environment~\cite{Kapral2013}. They are of interest for a number of reasons, including the potential to perform transport tasks~\cite{Patra2013}, and exhibit new emergent phenomena~\cite{MCMetal2013,Bech,Theurkauff2012,NYU2013,Bech2,Bricard2013}. A variety of subtly different methods, all based on the catalytic decomposition of dissolved fuel molecules, have been shown to produce autonomous motion, or swimming.  Many of the examples rely on the same catalytic reaction, the breakdown of hydrogen peroxide (H$_2$O$_2$) into water (H$_2$O) and oxygen (O$_2$) with metallic platinum (Pt) as catalyst. 

\begin{figure}[b]
\includegraphics[width=1.00\columnwidth]{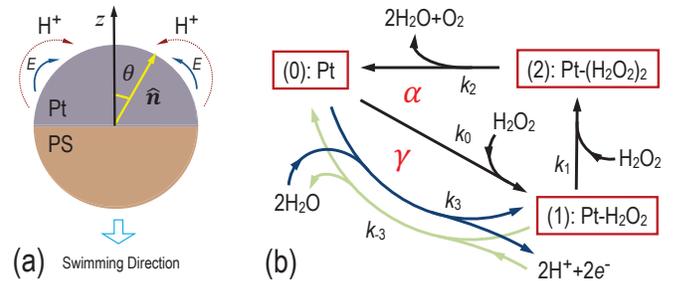}
\caption{(color online). (a) A half Pt-coated polystyrene (PS) Janus sphere showing the direction of flow of ions and the electric field. (b) The catalytic reaction scheme showing the competing reactions which have the topological structure of two coupled loops~\cite{Khudaish}. Loop $\alpha$ is the main nonequilibrium cycle that involves only uncharged species, and Loop $\gamma$ is linked to the production of charged intermediates H$^+$, $e^-$.
}
\label{fig:1}
\end{figure}

Commonly studied systems are catalytic bimetallic rod shaped devices~\cite{Kline2005} and non-conducting spherical Janus particles that are half-coated with catalyst~\cite{Howse2007} [see Fig. \ref{fig:1} (a)]. The propulsion mechanism is thought to be phoretic in nature~\cite{Anderson-review,Golestanian2007}, but the details remain the subject of debate. More and better experimental data are required to test the proposed mechanisms. For bimetallic swimmers, a plausible proposal is that the two metallic segments, usually platinum and gold, electrochemically reduce the dissolved fuel, in a process that results in electron transfer across the rod~\cite{Paxton2006}. This together with proton movement in the solution \cite{bachtold} {\em and} the interaction between the resulting self-generated electric field and the charge density on the rod produces (self-electrophoretic) motion~\cite{Moran2011}. The direction of travel and swimming speed for arbitrary pairs of metals are well understood in the context of this mechanism~\cite{Wang2006}, as well as the link between fuel concentration and velocity~\cite{Sabass2012}.

For Pt-insulator Janus particles, the absence of conduction between the two hemispheres suggests a mechanism independent of electrokinetics. Hence, a natural first proposal is that a self-generated gradient of product and reactants can lead to motion via self-diffusiophoresis \cite{Golestanian2005}, provided the colloid is sufficiently small \cite{Gibbs2009}. A number of predictions have been made based on this mechanism \cite{Golestanian2005,ruckner2007,udo,Valadares2010,popescu3,Brady,SharifiMood} which have to date shown good agreement with the experimental dependency of swimming velocity on the size of the colloid~\cite{Ebbens2012}, and fuel concentration~\cite{Howse2007}. It would thus appear that a key difference between the bimetallic and non-conducting Janus particles is that the motility in the latter system does not {\em require} conduction or electrostatic effects.

\begin{figure*}[t]
\includegraphics[width=1.00\linewidth]{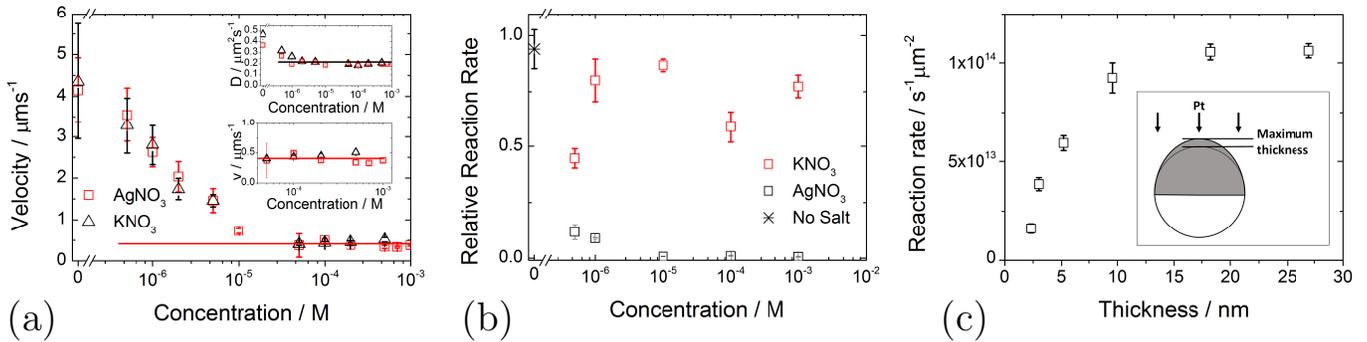}
\caption{(color online). (a) Effect of varying AgNO$_3$ and KNO$_3$ concentration on the propulsion velocity for 2 $\mu$m diameter Platinum Janus particles in 10 \% w/v H$_2$O$_2$ aqueous solutions. Top inset:  The corresponding translational diffusion coefficients as a function of salt concentration. The horizontal line indicates the expected diffusion coefficient for colloids of this size calculated using the Stokes-Einstein relation. Bottom inset: magnified view of the saturation limit of the velocity, with the average (solid line) being $v=0.44 \pm 0.02$ $\mu$m s$^{-1}$. (b) Relative initial H$_2$O$_2$ reaction rate as a function of salt concentration, measured on 1 cm$^2$ squares of 10 nm thick evaporated platinum.  The no-salt data point indicates a sample that was twice consecutively analysed in salt free hydrogen peroxide, and so acts as a control for the remainder of the data-points.
(c) Reaction rate as a function of average thickness of the Pt layer. Inset: a schematic view of the thickness profile of the Pt coating \cite{Campbell2013}.}
\label{fig:2}
\end{figure*}

Here we present an experimental and theoretical study which demonstrates however that electrokinetic effects \cite{ignacio} can also play a role in the motion of non-conducting spherical Janus particles. We find that their motion is due to a combination of neutral and ionic diffusiophoretic as well as electrophoretic effects whose interplay can be changed by varying the ionic properties of the fluid. This has great potential significance as the effect on the swimming behaviour, of solution properties such as temperature~\cite{Bala2009}, contaminants~\cite{Zhao2013}, pH, and salt concentration are of critical importance to potential applications that could eventually include drug delivery in the body~\cite{Patra2013}. Here, we focus in particular on the effect of salt-concentration on their swimming behaviour. For bimetallic nanorods, the addition of sub-millimolar amounts of salt can significantly reduce the swimming velocity~\cite{Paxton2006,Kagan2009}. This is consistent with self-electrophoresis~\cite{Paxton2006,Moran2011,Wang2012,Sabass2012}. However, experiments have also found that addition of some salts (e.g. AgNO$_3$) can lead to an {\em increase} in propulsion speed of bimetallic rods~\cite{Kagan2009}, which does not fit into a simple self-electrophoretic picture. This highlights the need for a better understanding of the mechanisms of Pt-catalytic self-propulsion. To study the salt dependency of the self-propulsion of Janus swimmers, we 
measure their velocity as a function of salt concentration as well as the reaction rates, to separate the observed slowing phenomena [shown in Fig. \ref{fig:2} (a)] from any effect of the salt on the platinum catalyst turnover rate.
To provide a comparison with the behaviour described above for bimetallic rods, we also incorporate AgNO$_3$, for which we observe no discernable change in the swimming velocity as compared with the degree of reduction for potassium salt [Fig. \ref{fig:2} (a)]---in sharp contrast to the case of bimetallic rods---together with an anomalous strong reduction in the overall reaction rate [Fig. \ref{fig:2} (b)]. Our accompanying theoretical analysis suggests that these effects could only be explained by a reaction scheme with {\em both} loops and charged intermediates in the catalytic reaction [see Fig. \ref{fig:1} (b)]. Furthermore, it also requires a systematic variation in the reaction rates across the Pt shell in the coated hemisphere, which could occur if the reaction rates have thickness dependence, as our Janus spheres have a systematic variation in the thickness of the coating (with a thicker coating at the pole than at the equator). We measure experimentally the variation of reaction rate with the thickness of the coating, and find indeed that it varies in the required fashion [Fig. \ref{fig:2} (c)].

\paragraph{Experimental.}
Janus particles were prepared by spin coating polystyrene spheres (Diameter = 2 $\mu$m, Duke Scientific) onto a glass slide, followed by thermal evaporation of a circa 10 nm thick coating of Platinum (Agar 99.99 \%).  These slides were re-suspended into a 10 \% w/v H$_2$O$_2$ solution, and videos (30 Hz) of the particles were captured using a PixelLink camera attached to a Nikon Eclipse microscope.  Labview Vision software was used to analyse these videos to find the $x$, $y$ centre for each particle in each frame, and calculate their Mean-Squared Displacement (MSD) as a function of time. As described elsewhere \cite{Howse2007,Dunderdale2012}, MSD vs time plots for the Janus swimmers that self-propel in a fixed direction normal to the platinum cap~\cite{showalter,Ebbens2011} can be used to extract the translational diffusion coefficient $D$, and the propulsion velocity $v$, as long as the time range fitted is short compared to the rotational constant. In the case of the 2 $\mu$m particles used here, this condition is satisfied by fitting to the first 0.5 seconds of data (theoretical rotational diffusion time=6.2 s). Appropriate solutions of KNO$_3$ and AgNO$_3$ (Sigma Aldrich 99.9 \%) were added to retain a constant peroxide concentration, and achieve salt concentrations in the range $10^{-7}$ to $10^{-3}$ M, and determination of propulsion velocity was repeated.
Reaction rate monitoring was performed using well-defined 1 cm$^2$ squares of platinum deposited on glass substrates at the required thickness.
The rate of reduction of hydrogen peroxide was monitored via UV absorption at 240 nm using a low volume auto sampling flow cell. For each condition reported, the initial reaction rate for the same platinum square was first determined in absence of the salt additive, and then again after the addition of salt to the desired concentration. This ensured any variation between the different platinum squares was minimised. The ratio of these two initial reaction rates was used to indicate any modification of reaction rate caused by the salt additive.

\paragraph{Results.}
Figure~\ref{fig:2} (a) shows the propulsion velocities and translational diffusion coefficients, determined using the MSD analysis as described above. It should be noted that this data is for colloids free to move in three dimensions in the bulk of the salt/peroxide aqueous solution, at least 100 $\mu$m away from the cuvette walls. The velocity observed for the salt free Janus swimmers is consistent with the magnitude we have reported elsewhere; small deviations between different batches of Janus swimmers can be assigned to issues with metal evaporation reproducibility and surface cleanliness. Figure~\ref{fig:2} (a) shows that the addition of both AgNO$_3$ and KNO$_3$ at concentrations in the range of $3 \times 10^{-7}$ M to $1 \times 10^{-5}$ M produces a monotonic reduction in propulsion velocity, showing a similar concentration dependence in both cases. Above salt concentrations of $3 \times 10^{-5}$ M, propulsion velocities reach a non-vanishing asymptotic value, corresponding to $v=0.44 \pm 0.02$ $\mu$m s$^{-1}$. This data shows that like the bimetallic nanorod swimmers described above, thin hemispherical shell, single-metal swimmers with a non-conductive, polymeric core are also susceptible to small amounts of salt. However, in contrast to the bimetallic rod data, the anomalous increase in propulsion velocity that was observed during the addition of AgNO$_3$ is not observed here. The inset of  Fig.~\ref{fig:2} (a) shows the diffusion coefficients that are also extracted during the MSD fitting procedure, which show good agreement with the value predicted by the Stokes-Einstein relation.
Figure~\ref{fig:2} (b)  summarises the relative reaction rate change induced by adding variable concentrations of the salts under consideration here. 
Although the measurement errors are large for the KNO$_3$ data, it appears that the addition of this salt in concentration ranges of $1 \times 10^{-6}$ to $1 \times 10^{-3}$M produces an average reduction in reaction rate to about 70 \% of the original value. The magnitude of this reduction is similar to that reported elsewhere for LiNO$_3$ and NaNO$_3$~\cite{Paxton2006}. The significance of the apparent non-monotonic dependence of the magnitude of the relative reaction rate is unclear, but some early literature on the effect of salts on the catalytic decomposition rate of H$_2$O$_2$ by colloidal Pt did report similarly complex behaviour~\cite{Heath2013}. 
It is clear, however, that the reduction in the propulsion velocity observed for KNO$_3$ concentrations cannot solely be explained by the reduction in the reaction rate. As for the AgNO$_3$ case, reaction rate data indicated that even micromolar additions produced a dramatic reduction in the reaction rate for the platinum square patch. Qualitatively, the reaction in these cases appeared to start more vigorously, but then significantly reduced within a few seconds of insertion of Pt into the AgNO$_3$ solution.
In fact, comparing the data point obtained at $1 \times 10^{-6}$ M with the corresponding swimmer velocity [Fig. \ref{fig:2}(a)] suggests a swimmer velocity significantly greater than that predicted by the reaction rate alone. We address this unexpected behaviour in our theoretical analysis below.

Figure~\ref{fig:2}(c) shows the absolute initial reaction rate for the decomposition of hydrogen peroxide by a series of flat glass substrates covered by evaporated platinum with varying thickness.
Atomic force microscopy of step-edges generated by masking some areas of the substrates with a mesh was used to accurately determine the thickness of each sample. It is evident that increasing the thickness of the platinum coating in the region 0-10 nm produces a strong variation in reaction rate. Sedimentation rate measurements reported for the Janus particles evaporated with 10 nm of platinum suggest an ellipsoidal shell deposition profile with maximum thickness at the pole, tapering to 0 nm at the equator~\cite{Campbell2013}. Consequently for such particles, a significant variation in reaction rate across their platinum cap can be expected.

\paragraph{Theory.}

A Janus sphere of radius $R$ has the catalytic reaction occurring on the Pt coated half. 
The normal to the plane splitting the hemispheres is chosen w.l.g. aligned in the $z$-direction [see Fig.~\ref{fig:1}(a)].
We propose a theoretical framework based on a detailed analysis of the reaction scheme for Pt catalysis of H$_2$O$_2$ degradation to water and O$_2$ based on the current electrochemical understanding~\cite{Khudaish} [see Fig.~\ref{fig:1}(b)]. A key point is that our analysis of self-propulsion takes account of the existence of charged intermediates within the catalytic reaction scheme  [see Fig.~\ref{fig:1}(b)].

The state of the system is described by the local state of the Pt on the coated hemisphere and the local concentrations of the various reactive species, H$_2$O$_2$, O$_2$ and H$^+$, denoted by $C_{hp} ({\bf x}), \; C_{o} ({\bf x}), \; C_{h} ({\bf x})$ respectively with bulk salt, H$^+$, and OH$^-$ concentrations denoted as $\bar C_s,\; \bar C_h, \; \bar C_{oh}$.
Positions in the bulk are denoted by $\bf x$ while positions on the surface are denoted by the unit vector $\nvec = \left( \sin\theta \cos \phi, \sin \theta \sin \phi, \cos \theta\right)$. The reaction scheme has three possible states of Pt [see Fig. \ref{fig:1} (b)]: free-Pt state (0), Pt-H$_2$O$_2$ state (1), Pt-(H$_2$O$_2)_2$ state (2), occupied with probability $p_0(\nvec),\; p_1(\nvec),\; p_2(\nvec)$ respectively. The rates  $k_i (\nvec)$ are taken to vary monotonically on the surface with a maximum at the poles and a minimum at the equator [see Fig. \ref{fig:2}(c)].
We consider the system in the steady state. Relations between the $p_i$'s and the rate constants are given by reaction flux conservation at the nodes of the reactions [see Fig.~\ref{fig:1}(b)].
The uncharged species diffuse freely in the solution, satisfying $D_{hp} \nabla^2 C_{hp} ({\bf x}) =0,
D_o \nabla^2 C_{o} ({\bf x}) =0$, while the charged species diffuse in an electric potential, $\Phi(\bfx)$: 
$
D_h \nabla \cdot \left[ \nabla C_{h} ({\bf x}) -  {e \beta C_{h}} \nabla \Phi  \right] =0$, where $\beta = (k_{\rm B} T)^{-1}$ and $e$ is the electronic charge. The boundary conditions on the surface are
$
-D_{hp} \partial_r C_{hp}|_R  = J_{hp} K (\nvec)$,
$-D_o \partial_r C_{o}|_R = J_o  K (\nvec)$,
$-D_h \left[\partial_r C_{h}  - \beta e C_{h} \partial_r \Phi\right] |_R= J_h K (\nvec) $,
where $K(\nvec)$ is a function that is 1 on the Pt hemisphere and 0 on the other, and $D_i$'s denote the respective diffusion constants. The diffusive currents on the surface due to the catalytic reaction are given by $J_{hp} = -k_0 C_{hp} p_0 - k_1 C_{hp} p_1$, $J_h=k_3 p_0 - k_{-3} C_h^2 p_1$, $J_o = k_2 p_2$. $\Phi$ obeys the Poisson-Nernst-Planck equation that takes the approximate form of $\nabla^2  \Phi ({\bf x}) =-{4 \pi e \over \epsilon} \left[   C_h (\bfx)+  \bar C_s e^{-e \beta \Phi (\bfx) } - (\bar C_s + \bar C_{oh})e^{+e \beta \Phi (\bfx)} \right]$ in stationary state. The fluid flow is considered in the Stokes (low Re) approximation. The problem is solved by a matched asymptotic boundary layer analysis~\cite{Anderson-review} separating the problem into a thin inner layer near the sphere surface and an outer charge-neutral region~\cite{Ibrahim2013}.
The variation in reaction rates from the equator to the poles, $k_i = k_i^{(0)} + \sum_l k_i^{(l)} P_l (\cos \theta)$,  leads to a self-generated electric field  in the outer region,  ${\bf E}  = - \nabla \Phi$ which can be expanded as a power series, $\beta e {\Phi}(r, \theta,\phi) = \sum_l A_l  (R/r)^{l+1}P_l  (\cos \theta)$ whose leading term is  $\Phi \simeq {A_1 \over \beta e}   \left( {R \over r}\right)^2 \cos \theta $ with
\beq
A_1 = { R \left [ k_2 C_{hp} \left(k_1 k_3 - k_0 k_{-3} \bar C_h^2  \right)\right]^{(1)}  \over \left [ 2 D_h (\bar C_s+\bar C_h) M +  (2 \bar C_s+\bar C_h)k_0 k_{-3} k_2 \bar C_h C_{hp} R  \right]^{(0)}} \nonumber
\eeq
where $M = k_0 k_1 C_{hp}^2 + \left(k_0 k_2 + k_1 k_3 + k_1 k_2 \right) C_{hp} + k_2 k_3 + k_2 k_{-3} \bar C_h^2$~\footnote{Any function of the rates $f(\{ k_i\})$ can be expressed w.l.g. as  $f(\{k_i\})^{(0)} + \sum_l  f  (\{k_i\})^{(l)} P_l (\cos \theta) $. 
}, and we have assumed $\bar C_{oh}=\bar C_{h}$. Note that the electric field ${\bf E}$ is not screened in the outer region, and that there is a corresponding long-range proton flux ${\bf J}_h$ around the sphere, with the two satisfying ${\bf J}_h=2 \beta e D_h (\bar C_s + \bar C_h) {\bf E}$.
The electric field has a nonzero tangential component in the outer region, as do the concentrations of the ionic species. Therefore, asymptotic matching of the inner and outer solutions leads to a slip velocity with both {\em electrophoretic} and {\em ionic diffusiophoretic} components ~\cite{Anderson-review}:
\(
\bfv_s^{ion} =  - {\epsilon \over 4 \pi \eta} \left[\zeta - \zeta^2 \frac{e\beta}{4} \left( {\bar C_s + \bar C_h \over \bar C_h}\right) \right]  ({\bf I} - \nvec\nvec) \cdot \nabla \Phi |_{r=R}
\)
(with $\zeta$ being the zeta-potential) in addition to the {\em diffusiophoretic} slip velocities $\bfv_s^{hp} = \mu_{hp}  ({\bf I} - \nvec\nvec) \cdot \nabla C_{hp}|_{r=R}, \bfv_s^{o} = \mu_{o}  ({\bf I} - \nvec\nvec) \cdot \nabla C_{o}|_{r=R}$.
Note that the coefficients $\mu_i$ depend on the range of the interaction of  the neutral species with the surface of the Janus particle~\cite{Anderson-review}.
Calculation of the fluid flow field gives rise to a swimming velocity with the diffusiophoretic components (as calculated in Ref. \cite{Ebbens2012}) and an additional (combined) ionic contribution 
\beq
{\bf U}_{ion} = \frac23 \; {A_1 k_{\rm B} T \over \eta R} \left(- {\sigma \over \kappa} + \frac18 {\sigma^2 \over \bar C_h} \right)\; \hat{\bf z},
\eeq
where $\sigma = {\epsilon \zeta \over 4 \pi e} \kappa$ is the surface charge density and $\kappa^{-1}$ is the Debye screening length defined by $\kappa^2 = 8 \pi \ell_{\rm B} (\bar C_s + \bar C_h)$, with $\ell_{\rm B}$ being the Bjerrum length. The forbidding expression above has a number of important simple features: {\bf (1)} it depends linearly on the fuel, $C_{hp}$ at low concentrations and saturates at high concentrations, {\bf (2)} it is independent of $R$ at small $R$ and behaves as $1/R$ for large $R$, and {\bf (3)} it is a monotonically decreasing function of salt concentration, $\bar C_s$ starting from a finite value when $\bar C_s=0$ and tending to zero as $\bar C_s$ becomes large. Hence at high salt concentration the swimming speed saturates to the neutral diffusiophoretic value, which corresponds to the plateau at $v=0.44$ $\mu$m s$^{-1}$ (reported above) in the current experiment. These features are reassuring, as they agree with all existing experimental results on this system \cite{Howse2007,Ebbens2012}.
The electrophoretic contribution, which can be much larger than the diffusiophoretic part, vanishes if there is no variation in the rates $k_i$ on the surface.
We note also that due to the existence of the two separate reaction loops, the overall catalytic reaction rate (measured from the current $J_o$ above) can be significantly reduced with only small reductions to the swimming speed; say by a significant decrease in $k_1$ \cite{Ibrahim2013}. This type of behaviour would be expected from any reaction scheme which has this topological structure.

\begin{acknowledgments}
This work was supported by EPSRC grants EP/J002402/1 (SE), EP/G04077X/1 (JRH), EP/G026440/1 (TBL, YI), and Human Frontier Science Program (HFSP) grant RGP0061/2013 (RG).
\end{acknowledgments}

\end{document}